# Giant oscillatory Gilbert damping in superconductor/ferromagnet/superconductor junctions

**Authors**

Yunyan Yao[1,2,†], Ranran Cai[1,2,†], Tao Yu[3], Yang Ma[1,2], Wenyu Xing[1,2], Yuan Ji[1,2], Xin-Cheng Xie[1,2,4,5], See-Hun Yang[6]*, and Wei Han[1,2]*

**Affiliations**

[1]International Center for Quantum Materials, School of Physics, Peking University, Beijing 100871, China.

[2]Collaborative Innovation Center of Quantum Matter, Beijing 100871, China.

[3]Max Planck Institute for the Structure and Dynamics of Matter, 22761 Hamburg, Germany

[4]CAS Center for Excellence in Topological Quantum Computation, University of Chinese Academy of Sciences, Beijing 100190, P. R. China

[5]Beijing Academy of Quantum Information Sciences, Beijing 100193, P. R. China

[6]IBM Research - Almaden, San Jose, California 95120, USA

[†]These authors contributed equally to the work

*Correspondence to: weihan@pku.edu.cn (W.H.); seeyang@us.ibm.com (S.H.Y.).

**Abstract**

Interfaces between materials with differently ordered phases present unique opportunities for exotic physical properties, especially the interplay between ferromagnetism and superconductivity in the ferromagnet/superconductor heterostructures. The investigation of *zero*- and $\pi$-junctions has been of particular interest for both fundamental physical science and emerging technologies. Here, we report the experimental observation of giant oscillatory Gilbert damping in the superconducting Nb/NiFe/Nb junctions with respect to the NiFe thickness. This observation suggests an unconventional spin pumping and relaxation *via* zero-energy Andreev bound states that exist *only* in the Nb/NiFe/Nb $\pi$-junctions, but not in the Nb/NiFe/Nb *zero*-junctions. Our findings could be important for further exploring the exotic physical properties of ferromagnet/superconductor heterostructures, and potential applications of ferromagnet $\pi$-junctions in quantum computing, such as half-quantum flux qubits.

**One sentence summary:** Giant oscillatory Gilbert damping is observed in superconductor/ferromagnet/superconductor junctions with varying the ferromagnet thickness.

## Introduction

The interplay between ferromagnetism and superconductivity has induced many exotic and exciting physical properties in ferromagnet (FM)/superconductor (SC) heterostructures (*1-3*). Of particular interest is the unconventional $\pi$-phase ground state SC/FM/SC junction that might be realized for certain FM thicknesses arising from the quantum intermixing of the wave functions between spin-singlet Cooper pairs in SC and spin-polarized electrons in FM (*1, 3, 4*). At the FM/SC interface, a Cooper pair moving into the FM will have a finite center-of-mass momentum, resulting in the oscillation of the real part of superconducting order parameter (Re $\{\Psi\}$) with respect to the FM thickness (Fig. 1A) (*1, 5, 6*). Depending on the FM thicknesses, the Cooper pair wavefunctions in the two superconductors on either side of the FM can have a phase difference from zero or $\pi$, forming so-called *zero*-junctions with positive Josephson coupling (Fig. 1B) or $\pi$-junctions with the negative Josephson coupling (Fig. 1C). The FM $\pi$-junctions can be used for quantum computing applications (*7, 8*), as half quantum flux qubits (*9*). Due to the scientific and technical importance, the research on the FM $\pi$-junctions has been active for the last two decades (*6, 10-13*). Previous experimental studies have demonstrated the switching between *zero*- and $\pi$-junctions in SC/FM/SC structures by varying the temperature and the FM thickness (*11, 14-17*). These reports mainly focus on the electrical properties of the FM *zero*- and $\pi$-junctions. Recently, dynamic spin injection into SCs has attracted considerable interest both the experimentally (*18-21*) and theoretically (*22-26*). However, the spin-dependent properties in FM *zero*- and $\pi$-junctions have not been explored yet. The investigation of the spin-dependent properties requires the spin current probes, such as the dynamical spin pumping (*27*). Furthemore, for the application of the FM $\pi$-junctions in quantum computing technologies (*9*), the magnetization/spin dynamic properties are extremely important to be studied.

Here, we report the experimental observation of giant oscillatory Gilbert damping in the superconducting Nb/NiFe/Nb junctions with respect to the NiFe thickness, which can be qualitatively explained by the different spin pumping efficiency via the Andreev bound states (ABS) of Nb/NiFe/Nb *zero-* and *π-*junctions. Using a minimal model based on the ABS, we show that an unconventional spin pumping into the zero-energy ABS penetrated into SCs could occurs only for the *π*-junctions, which can lead to the oscillatory Gilbert damping as a function of the NiFe thickness.

## Results

Figures 1D and 1E show the schematics of the spin pumping, magnetization dynamics, and enhanced Gilbert damping in the SC/FM/SC *zero-* and *π*-junctions. Spin pumping refers to the spin-polarized current injection to non-magnetic materials from a FM with precessing magnetization around its ferromagnetic resonance (FMR) conditions (*28, 29*). In FM and its heterostructures, the Gilbert damping ($\alpha$) characterizes the magnetization dynamics, as described by the Landau-Lifshitz-Gilbert formula with an additional Slonczewski-torque term (*30-32*):

$$\frac{d\boldsymbol{m}}{dt} = -\gamma \boldsymbol{m} \times \boldsymbol{H}_{eff} + \alpha \boldsymbol{m} \times \frac{d\boldsymbol{m}}{dt} + \frac{\gamma}{M_s V}\left(\frac{\hbar}{4\pi} g^{\uparrow\downarrow} \boldsymbol{m} \times \frac{d\boldsymbol{m}}{dt}\right) \qquad (1)$$

where $\boldsymbol{m} = \boldsymbol{M}/|\boldsymbol{M}|$ is the magnetization unit vector, $\gamma$ is the gyromagnetic ratio, $\boldsymbol{H}_{eff}$ is the total effective magnetic field, $M_s = |\boldsymbol{M}|$ is the saturation magnetization, and $g^{\uparrow\downarrow}$ is the interface spin mixing conductance. The pumped spin current from FM into SCs can be expressed by $\mathbf{J_s} = \frac{\hbar}{4\pi} g^{\uparrow\downarrow} \boldsymbol{m} \times \frac{d\boldsymbol{m}}{dt}$ (*29*). The spin pumping into the SCs give rise to an enhanced Gilbert damping constant that is proportional to the spin pumping current ($\alpha_{sp} \sim J_s$) (*29*). Fig. 1E illustrates the pumped spin current mediated by the zero-energy ABS inside the superconducting gap in *π*-junctions, which will be discussed later in details. While for a *zero*-junction, the pumped spin current is mediated by the ABS near the superconducting gap (Fig. 1D). The ABS can be formed within the FM layer and then extended into the interface of SCs with the superconducting coherent length scale (*33, 34*).

The SC/FM/SC junctions consist of a NiFe (Ni$_{80}$Fe$_{20}$) layer (thickness: ~ 5 - 20 nm) sandwiched by two Nb layers (thickness: 100 nm) grown by magnetron sputtering (see Methods and fig. S1). To maximize the integrity of samples for a systematic study, more than tens of samples are grown in each run via rotation mask technique in a sputtering system, which is the same as in the previous study of the oscillatory exchange coupling in magnetic multilayer structures (*35*). The Gilbert damping and spin pumping are measured by the ferromagnetic resonance (FMR) technique (see Methods for details).

Above the $T_C$ of Nb, spin pumping in the Nb/NiFe/Nb junctions leads to the spin accumulation in Nb near the interface, which can be described by the spin-dependent chemical potentials, as illustrated in Fig. 2A. The Gilbert damping of NiFe in the Nb/NiFe/Nb junctions is determined from the microwave frequency-dependent FMR spectra (fig. S2). A typical FMR curve with the Lorentzian fitting is shown in Fig. 2B, from which the half linewidth (ΔH) can be obtained. The Gilbert damping can be extracted from the best linear-fitting curve of ΔH vs. *f* (Fig. 2C). Figure 2D shows the NiFe thickness dependence of the Gilbert damping in the Nb/NiFe/Nb junctions measured at *T* = 10, 15, and 20 K, respectively. Interestingly, an oscillating feature of the Gilbert damping is observed as a function of $d_{\text{NiFe}}$ in the region of $d_{\text{NiFe}} < $ ~15 nm. This oscillating behavior can be attributed to the quantum-interference effect of angular momentum transfer between the local precessing magnetic moment and conduction electrons in thin NiFe that was theoretically predicted by Mills (*36*), but has not been experimentally reported yet. Above $T_C$, the continuous energy bands of Nb, similar to the normal metal in the Mills theory, overlap with both spin-up and spin-down bands of NiFe at the interface, thus allowing the conducting electrons in NiFe to flip between the spin-down and spin-up states. As illustrated in the inset of Fig. 2D, one spin-down electron scatters with the local magnetic moment and then flips to the spin-up polarization, giving rise to the angular momentum transfer between the spin-polarized electrons and the magnetic moment. Besides the change of angular momentum, the momentum of the electron also changes (Δ$k$), due to different Fermi vectors for spin-up ($k_{F\uparrow}$) and spin-down ($k_{F\downarrow}$) electrons with exchange splitting (Fig. 2A). When the NiFe layer is thin enough to become comparable with $\frac{1}{\Delta k}$, quantum-interference effect of the spin-polarized electrons shows up, which gives rise to the oscillating spin-transfer torque to the NiFe. When the NiFe thickness is $2n\pi/[k_{F\uparrow} - k_{F\downarrow}]$ (*n* is an integer), the matching of the quantum levels between the spin-up and spin-down electrons in NiFe induces smaller Gilbert damping. On the other hand, when the NiFe thickness is $(2n+1)\pi/[k_{F\uparrow} - k_{F\downarrow}]$, a larger Gilbert damping is induced. Consequently, the Gilbert damping in the Nb/NiFe/Nb structures oscillates with a period of $2\pi/[k_{F\uparrow} - k_{F\downarrow}]$ (Supplementary Materials S1). Experimentally, an oscillating period (*λ*) of ~ 1.8 nm is identified (see the red dashed arrow in Fig. 2D). At $T = 50$ K, the oscillating feature disappears since the quantum-interference effect is smeared by thermal excitations (fig. S3).

Next, we investigate the spin pumping and spin transfer torque of the Nb/NiFe/Nb junctions in the superconducting states below $T_C$ with a superconducting gap (Fig. 3A). $T_C$ in the Nb/NiFe/Nb junctions is obtained from typical four-probe resistance measurement as a function of the temperature. A typical temperature-dependent resistance curve measured on the Nb/NiFe (12 nm)/Nb junction is shown in Fig. 3b, indicating the $T_C$ of ~ 8.6 K. As $d_{NiFe}$ changes, $T_C$ of the Nb/NiFe/Nb junctions exhibits little variation between ~ 8.4 and ~ 8.9 K (fig. S4). Similar to the normal states of Nb, the Gilbert damping below $T_C$ is also obtained from the best linear-fitting result of the half linewidth vs. frequency (fig. S5). During the FMR measurement, $T_C$ varies a little (< 1 K) (Fig. S6). As the temperature decreases, $\alpha$ decreases abruptly from ~ 0.012 to ~ 0.0036 across the $T_C$ (fig. 3C), which indicates the decrease of spin current injected into Nb due to the formation of superconducting gap below $T_C$. This observation is consistent with previous reports on spin pumping into SCs where the spin current is mediated by Bogoliubov quasiparticles (*18, 19, 37*). As the temperature decreases far below the $T_C$, the quasiparticle population dramatically decreases, leading to reduced spin pumping and Gilbert damping.

Remarkably, the oscillating amplitude of the Gilbert damping of the Nb/NiFe/Nb junctions as a function of the NiFe thickness is dramatically enhanced as the temperature decreases into the superconducting states of Nb (Fig. 3D). At $T = 4$ K, the oscillating magnitude of the Gilbert damping constant is ~ 0.005 for the first three oscillations, which is comparable to the background value of ~ 0.006. The obtained Gilbert damping values are not affected by thermal cycles, and the large oscillating feature has been confirmed on a different set of samples. Such a giant oscillation of the Gilbert damping cannot be explained by spin pumping of Bogoliubov quasiparticle-mediated spin current in SCs. Since as the temperature decreases, the population of the Bogoliubov quasiparticles monotonically and rapidly decreases with an increase of the SC gap, which would lead to lower Gilbert damping and also smaller oscillation compared to the normal states. Note that the oscillating period of the Gilbert damping at $T = 4$ K is the same as that at $T = 10$ K that is supposed to be $2\pi/[k_{F\uparrow} - k_{F\downarrow}]$ due to the quantum interference effect. Such oscillating period of $2\pi/[k_{F\uparrow} - k_{F\downarrow}]$ is the also same as that of the *zero*- and $\pi$-phase ground states transitions in FM Josephson devices, which is equal to the coherence length in NiFe film of $2\pi/[k_{F\uparrow} - k_{F\downarrow}]$ in the ballistic regime (*1, 11, 17*), and $\sqrt{\hbar D_{diff} / E_{ex}}$ in the diffusive regime ($D_{diff}$ is the diffusion coefficient, and $E_{ex}$ is the exchange energy). The observed oscillating period of ~ 1.8 nm in our study is similar to the *zero - π* oscillating period measured in the NiFe Josephson junctions in the diffusive regime reported previously (*11, 17*).

The Gilbert damping difference ($\Delta\alpha$) between the *zero-* and *π*-junctions is extracted as a function of NiFe thickness, as shown in Fig. 4A. We assume the larger Gilbert damping for the *π*-junctions and smaller values for the *zero*-junctions, which will be discussed later in details. The thickness-dependent Gilbert damping of the *zero-* and *π*-junctions are expected to both behave as $\alpha \sim 1/d_{NiFe}$ (*29*). Hence, we can treat them separately, as illustrated by the guide lines in the inset of Fig. 4A, and $\Delta\alpha$ is obtained by subtracting the fitted 1/d curve for the expected *zero*-junctions (black dashed line). Clearly, there is a pronounced oscillating feature of $\Delta\alpha$ for the Nb/NiFe/Nb junctions with NiFe thickness from ~ 5 nm to ~ 11 nm. When the NiFe thickness is above ~ 11 nm, the oscillating feature of the Gilbert damping is largely suppressed compared to thinner NiFe junctions. This feature might be associated with the strong Josephson coupling for thin NiFe junctions and the exponential decaying of the Josephson coupling as the NiFe thickness increases (*11, 17*). To confirm this, the Josephson junctions are fabricated using the shadow mask technique, and a Josephson coupling is observed from the Nb/NiFe (5 nm and 10 nm)/Nb junctions (Supplementary Materials and fig. S7).

## Discussion

Let us discuss the physical mechanism that induces the giant oscillating Gilbert damping in the following. Apart from the spin pumping via ABS discussed above (Fig. 1D and 1E), the spin current in SCs can also be mediated by Bogoliubov quasiparticles (fig. S8A) (*18, 19, 22, 23, 38*), spin-triplet pairs (fig. S8B) (*3*). Regarding Bogoliubov quasiparticles, they populate around the edge of superconducting gap at elevated temperatures close to $T_C$ (*39*). As shown both theoretical and experimental studies, the enhanced Gilbert damping in the SC/FM/SC heterostructures happens around $T_C$ (*18, 19, 22, 23, 38*). As the temperature decreases down to 0.5 $T_C$, the Bogoliubov quasiparticles are mostly frozen out, for which the spin pumping is forbidden that will no longer contribute to the enhanced Gilbert damping. Hence, the Bogoliubov quasiparticles are very unlikely to account for our experimental results. Regarding the spin-triplet pairs, it has been shown in previous studies that the spin-triplet current under FMR conditions and spin triplet correlations would be different for *zero*- and $\pi$-junctions (*4, 38, 40, 41*), which might result in different Gilbert damping theoretically. However, in our study, there are not spin sinks adjacent to the Nb layers, thus not allowing the spin-triplet Cooper pairs to be relaxed in the Nb. This is different from previous report on the Pt/SC/FM/SC/Pt heterostructures(*20*), where the Pt is used as the spin sink. Experimentally, as the temperature below $T_C$, the Gilbert damping exhibits a monotonic decrease for the Nb/NiFe/Nb heterostructures (Fig. 3C), which is different from the enhanced Gilbert damping due to spin-triplet pairs (*20*). Furthermore, no Josephson current in the Nb/NiFe/Nb heterostructures is observed in Nb/NiFe (30 nm)/Nb junction (fig. S7), which indicates the absence of long-range spin-triplet Josephson coupling. Both these experimental results indicate that the contribution from the spin-triplet pairs is not significant to the enhanced Gilbert damping in the superconducting Nb/NiFe/Nb junctions.

To our best understanding, the most reasonable mechanism is the spin pumping via the ABS, which can qualitatively describe our experimental observation. Previous studies have demonstrated that the energy of ABS inside the superconducting gap depends on the superconducting-phase (*42, 43*). For the FMR measurement under open-circuited conditions, the inversion symmetry of the current-phase ($\varphi$) relationships is preserved (*43-45*). For $\pi$-junctions, there is a $\pi$-phase shift in the current-phase relationship curves compared to *zero*-junctions, i.e., the properties of $\varphi = 0$ of a $\pi$-junction is the same as those of $\varphi = \pi$ of a *zero*-junction. Since this $\pi$-phase shift is already taken into account by the FM exchange field, the ABS energy of the $\pi$-junctions can be obtained at $\varphi = 0$ in the ground states, which is similar to that of $\varphi = \pi$ of *zero*-junctions.

For $\pi$-junctions, ABS is located around the zero-energy inside of the superconducting gap (Fig. 1D). The ABS could penetrate into the superconducting Nb films with scale of superconducting coherent length (~ 30 nm), which is evanescent to dissipate the spin angular momentum (*25, 26, 44*). As shown in Fig. 4B, the transfer efficiency of spin angular momentum via the zero-energy ABS can lead to an enhanced Gilbert damping. Whileas, for *zero*-junctions, the distribution of the ABS is near the edge of the superconducting gap (Fig. 1C), thus, the spin pumping efficiency is suppressed due to the reduced population of the ABS at low temperatures (Fig. 4C). Furthermore, the oscillatory energy levels of the ABS between the *zero*- and $\pi$-junctions is also consistent with the density of states (DOS) oscillating in superconductors between the *zero*- and $\pi$-junctions (*1, 6,*

*38, 44, 46*). In consequence, as the NiFe thickness increases, the oscillatory spin pumping efficiency via ABS at the FM/SC interface (or DOS in SCs) gives rise to the oscillatory Gilbert damping. We have proposed a simplified model for the case of ideal transparency of electrons (Supplementary Materials S2 and fig. S9). For the less transparency cases, i.e., in diffusive regime, (*42, 43*), the energy levels of the ABS in $\pi$-junctions locates away from zero-energy, but they are still much smaller than those of the ABS in *zero*-junctions. Actually, the similar oscillating behaviors of ABS (or DOS) can be preserved in the diffusive regime (*6, 46*). Hence, an oscillating spin pumping efficiency would also be expected in the diffusive regime, which could lead to the oscillating Gilbert damping observed in our experiment. To fully understand the experimental observation of the oscillatory Gilbert damping and the detailed spin relaxation process in the diffusive regime, further theoretical studies are needed.

Furthermore, the control samples of bilayer Nb/NiFe heterostructures do not exhibit the large oscillatory feature for the Gilbert damping as the NiFe thickness varies at $T = 4$ K (fig. S10), which further presents the important role of phase difference across NiFe in the large the oscillatory Gilbert damping observed in the trilayer Nb/NiFe/Nb heterostructures.

In conclusion, giant oscillatory Gilbert damping is observed in the superconducting Nb/NiFe/Nb junctions with respect to the NiFe thickness. To our best knowledge, neither the Bogoliubov quasiparticles, nor the spin-triplet pairs are relevant to this observation. The most possible explanation for such giant oscillatory Gilbert damping could be related to the different ABS energy levels and the DOS at the NiFe/SC interface in *zero*- and $\pi$- junctions. To fully understand these results, further theoretical studies are needed. Looking forward, our experimental results might pave the way for controlling the magnetization dynamics by the superconducting phase in a FM Josephson junction in the SQUID setup, and could be important potential applications of ferromagnet $\pi$-junctions in quantum computing, such as half-quantum flux qubits.

## Materials and Methods

### Materials growth

The SC/FM/SC heterostructures consisting of Nb (100 nm) and $Ni_{80}Fe_{20}$ (NiFe; ~ 5 - 20 nm) were grown on thermally oxidized Si substrates in a d.c. magnetron sputtering system with a base pressure of $\sim 1\times 10^{-8}$ torr. To systematically vary the NiFe thickness that is crucial for the quantum-size effect, we adopted the rotating multi-platter technique that allows us to grow dozens of Nb/NiFe/Nb samples in each run (*35*). The thickness of the Nb layer is fixed to be ~100 nm that is much larger than the spin diffusion length of Nb (*20, 47*). After the growth, a thin $Al_2O_3$ layer (~ 10 nm) was deposited *in situ* as a capping layer to avoid sample degradation against air/water exposure. The crystalline properties of Nb/NiFe/Nb heterostructures were characterized by X-ray diffraction (fig. S1A) and high-resolution cross-sectional transmission electron microscopy (fig. S1B) using a 200-kV JEOL 2010F field-emission microscope. The NiFe thickness is determined by the growth rate that is calibrated by TEM measurement, where the uncertainty of the NiFe thickness is obtained to be smaller than ~ 0.8 nm (fig. S1B). The resistivity of the NiFe layers (thickness: 5 - 20 nm) is ranging from 60 to 35 μΩ·cm, which corresponds to the mean free path between 2.3 and 3.9 nm.

**<u>Ferromagnetic resonance measurement.</u>**

The spin pumping of Nb/NiFe/Nb heterostructures was characterized via FMR using the coplanar wave guide technique connected with a vector network analyzer (VNA; Agilent E5071C) in the variable temperature insert of a Physical Properties Measurement System (PPMS; Quantum Design) (*19*). The FMR spectra were characterized by measuring the amplitudes of forward complex transmission coefficients ($S_{21}$) as the in-plane magnetic field decreases from 4000 to 0 Oe under the microwave power of 1 mW. The typical FMR results measured on the Nb/NiFe (12 nm)/Nb heterostructures are shown in the fig. S2A (*T* = 10 K) and fig. S5A (*T* = 4 K). Weaker FMR signals are observed in the superconducting states compared to the normal states.

The half linewidth (Δ*H*) can be obtained by the Lorentz fitting of the magnetic field-dependent FMR signal following the relationship (figs. S2B and S4B):

$$S_{21} \propto S_0 \frac{(\Delta H)^2}{(\Delta H)^2 + (\boldsymbol{H} - \boldsymbol{H_{res}})^2} \tag{3}$$

where $S_0$ is the coefficient for the transmitted microwave power, ***H*** is the external in-plane magnetic field, and $\boldsymbol{H_{res}}$ is the resonance magnetic field. The Gilbert damping constant (α) can be obtained from the slope of the best linear-fitting results of the Δ*H* vs. the microwave frequency (*f*) (*48-51*):

$$\Delta H = \Delta H_0 + \left(\frac{2\pi\alpha}{\gamma}\right) f \qquad (4)$$

where $\Delta H_0$ is the zero-frequency line broadening that is related to the inhomogeneous properties, and $\gamma$ is the gyromagnetic ratio. From the best linearly fits of the $\Delta H$ vs. f results measured on the typical Nb/Py (12 nm)/Nb sample (red lines in figs. S2C and S5C), $\alpha$ is determined to be 0.012 and 0.0054 at T = 10 and 4 K, respectively. A larger zero-frequency line broadening $\Delta H_0$ is observed for the superconducting state compared to the normal state of Nb/Py/Nb heterostructures, which could be attributed to Meissner screening effect and the formation of trapped magnetic fluxes in Nb (*51*). The thickness dependent $\Delta H_0$ is shown in fig. S11C, and no obviously oscillatory behaviors are observed.

The effective magnetization and the gyromagnetic ratio can be fitted via the in-plane Kittel formula (*51*):

$$f_{res} = \frac{\gamma}{2\pi}\sqrt{(H_{res} + h)(H_{res} + h + 4\pi M_{eff})}, \qquad (5)$$

where $f_{res}$ and $H_{res}$ are the resonant microwave frequency and magnetic field respectively, $4\pi M_{eff}$ is the effective saturated magnetization, and $h$ is the shifted magnetic field induced by superconducting proximity effect. The thickness-dependent gyromagnetic ratio and effective magnetization can be found in fig. S11A and fig. S11B. Both parameters do not exhibit any oscillatory features as the Gilbert damping does (Fig. 4A), which demonstrates that the oscillatory Gilbert damping is not caused by any unintentional experimental error.

**<u>Superconducting transition temperature measurement.</u>**

The superconducting transition temperature (*T<sub>C</sub>*) of the Nb/NiFe/Nb heterostructures was determined via the zero-resistance temperature measured by four-probe method in a PPMS using standard a.c. lock-in technique at a low frequency of 7 Hz. The $T_C$ of Nb (100 nm)/NiFe/Nb (100 nm) heterostructures exhibits little variation as a function of the NiFe thickness (fig. S4). It is noticed that the FMR measurement can affect the $T_C$ a little (< 1 K), as shown in fig. S6.

**Supplementary Materials**

Supplementary Materials and Methods

fig. S1. The crystalline properties of the Nb/NiFe/Nb heterostructures.

fig. S2. Gilbert damping measurement of Nb/NiFe/Nb heterostructures at $T = 10$ K.

fig. S3. NiFe thickness dependence of Gilbert damping at T = 50 K.

fig. S4. NiFe thickness dependence of $T_C$ for the Nb/NiFe/Nb heterostructures.

fig. S5. Measurement of the Gilbert damping of Nb/NiFe/Nb heterostructures at $T = 4$ K.

fig. S6. The effect of FMR measurement on the $T_C$ of Nb/NiFe/Nb heterostructures.

fig. S7. The measurement of Josephson coupling in Nb/NiFe/Nb junctions.

fig. S8. Illustration of magnetization dynamics and spin pumping in the SC/FM/SC heterostructures due to Bogoliubov quasiparticles and equal spin-triplet Cooper pairs.

fig. S9. Calculation of the enhanced Gilbert damping due to spin pumping via the ABS at $T = 4$ K.

fig. S10. Gilbert damping of control sample of bilayer Nb/NiFe junctions.

fig. S11. Thickness dependence of gyromagnetic ratio, effective magnetization and inhomogeneous half-linewidth.

**Acknowledgments**

**General**: We acknowledge the fruitful discussion with Sadamichi Maekawa, Ziqiang Qiu, Zhe Yuan, Ke Xia, Young Sun, and Kei Yamamoto. Y.Y., R.C., Y.M., W.X., Y.J., X.C.X., and W.H. acknowledge the financial support from National Basic Research Programs of China (No. 2019YFA0308401), National Natural Science Foundation of China (No. 11974025), Beijing Natural Science Foundation (No. 1192009), and the Strategic Priority Research Program of the Chinese Academy of Sciences (No. XDB28000000). T.Y. is financially supported by DFG Emmy Noether program (SE 2558/2-1).

**Author contributions:** W.H. conceived and supervised the project. Y.Y. and R.C. performed the ferromagnetic resonance measurements. Y.Y. and Y.M. performed X-ray diffraction measurements. T.Y. performed the theoretical calculations. S.H.Y. synthesized the Nb/NiFe/Nb heterostructures. Y.Y. and W.H. wrote the manuscript with the contribution from all authors. All the authors discussed the results.

**Competing interests:** The authors declare no competing interests.

**Data Availability:** All data needed to evaluate the conclusions in the paper are present in the paper and/or the Supplementary Materials.

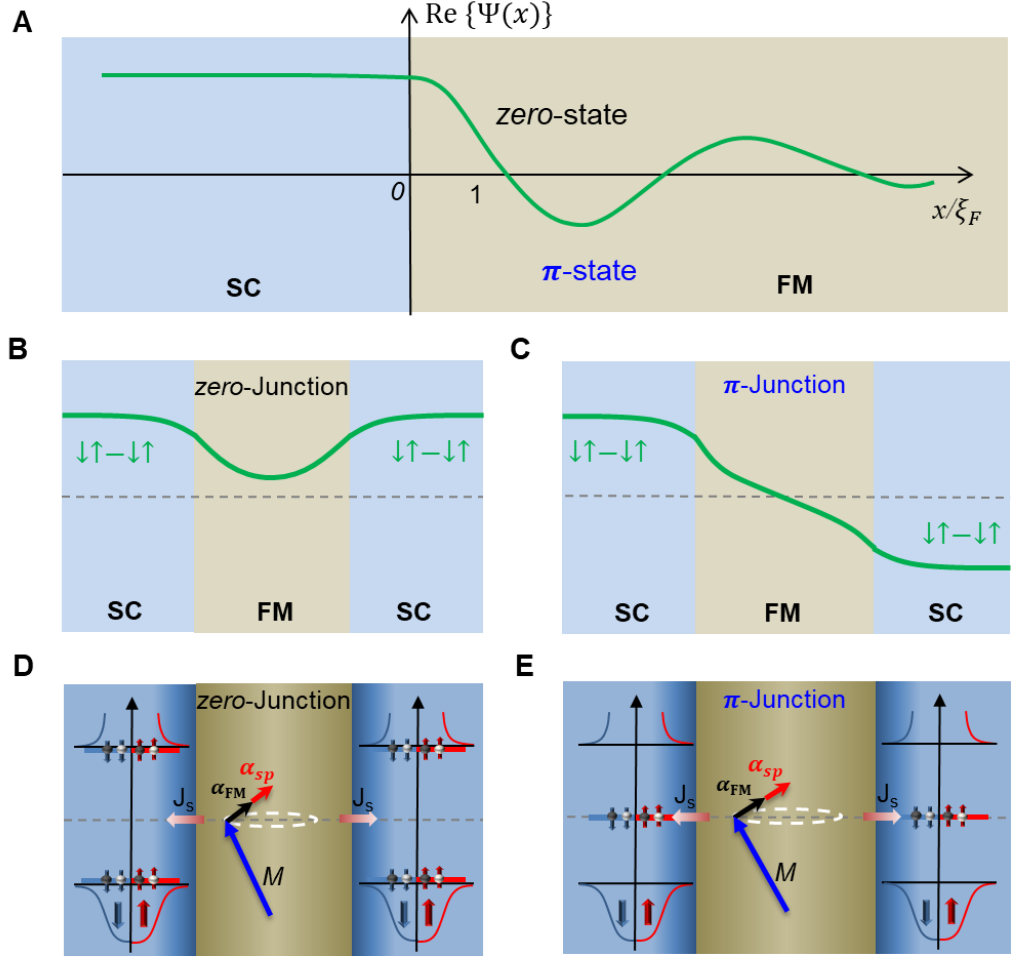

**Fig. 1. Illustration of magnetization dynamics and spin pumping in the SC/FM/SC heterostructures.** (**A**) The oscillatory real part of the superconducting order parameter (Re{Ψ}, green curve) penetrated into FM leads to the *zero*-state and π-state. (**B**) The symmetric order parameter in the *zero*-junctions. (**C**) The anti-symmetric order parameter in the π-junctions. (D-E) Spin pumping via the ABS in SCs in the *zero*- and π-junctions. $M$ and $\alpha_{FM}$ are the magnetization and Gilbert damping of the FM layer itself, and $\alpha_{sp}$ is the enhanced Gilbert damping, which arises from the spin dissipation in SC layers during the spin pumping process.

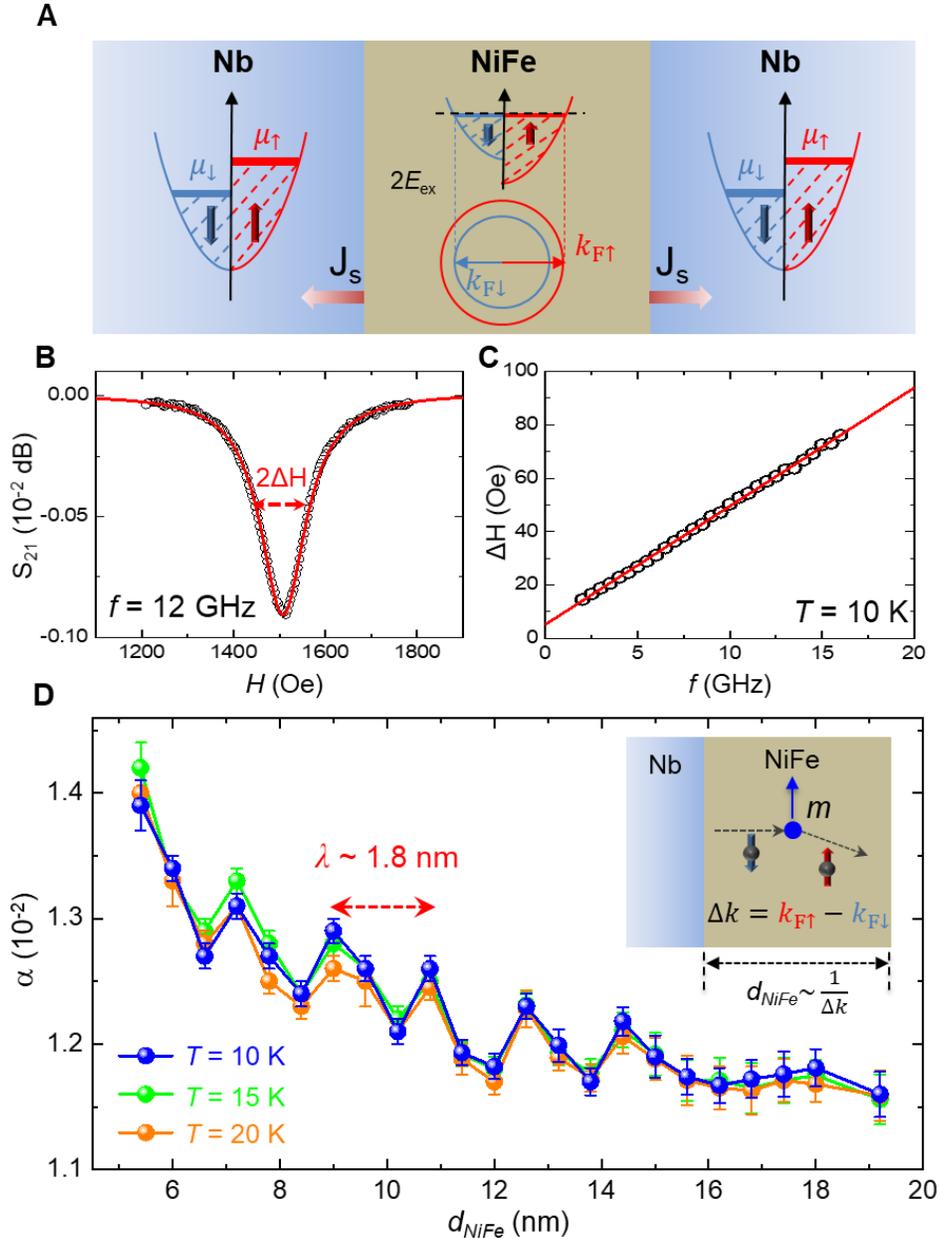

**Fig. 2. Oscillatory Gilbert damping of the Nb/NiFe/Nb heterostructures above $T_C$.** (**A**) The illustration of spin pumping into the normal states of Nb layers and the electronic band structure of NiFe with different spin-up and spin-down Fermi vectors ($k_{F\uparrow}$ and $k_{F\downarrow}$) due to the exchange splitting ($2E_{ex}$). The spin pumping gives rise to the spin accumulation in the Nb layers, indicated by the spin-split chemical potential ($\mu_\uparrow$ and $\mu_\downarrow$). (**B**) A typical FMR curve measured with $f = 12$ GHz (black circles) and the Lorentzian fitting curve (red line) measured on Nb/Py (12 nm)/Nb. ΔH is the half linewidth at the half maximum of FMR signal. (**C**) The determination of the Gilbert

damping from ΔH vs. *f*. The red line represents the best linear-fitting curve. (**D**) The oscillatory Gilbert damping as a function of NiFe thickness ($d_{NiFe}$) measured at $T = $ 10, 15, and 20 K, respectively. The experimental oscillating period ($\lambda$) is marked by the red dashed arrow. The inset: Illustration of the quantum-interference effect of the angular momentum transfer between the local magnetic moment and the spin-polarized electrons. When the NiFe thickness decreases to scale of $\frac{1}{\Delta k}$, the quantum-interference effect starts to be significant in the angular momentum transfer and spin pumping into the Nb layers.

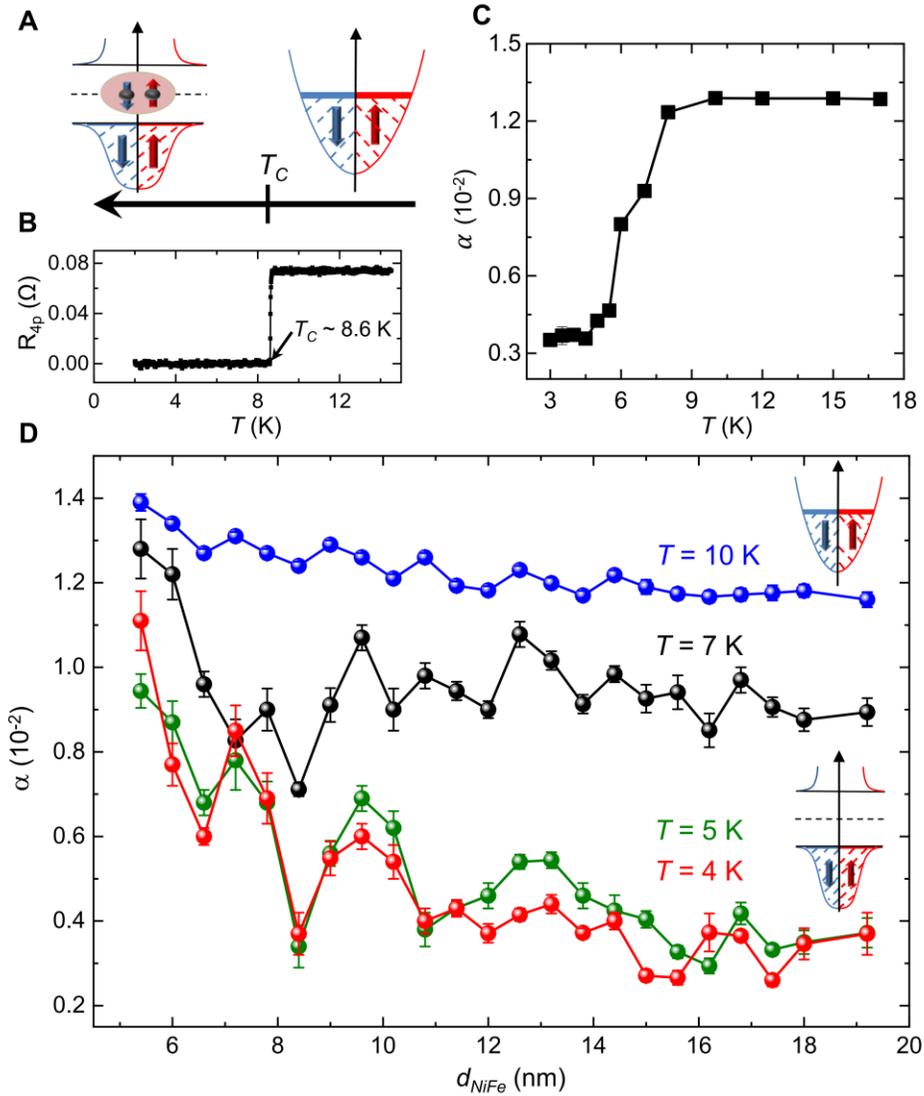

**Fig. 3. Giant oscillatory Gilbert damping in Nb/NiFe/Nb heterostructures below $T_C$.** (**A**) The illustration of electronic band structures of Nb in the normal and superconducting states. (**B**) The determination of $T_C$ via zero-resistance temperature measured on the typical Nb/NiFe (12 nm)/Nb heterostructures. (**C**) Temperature dependence of Gilbert damping of the typical Nb/NiFe (12 nm)/Nb heterostructures. (**D**) The oscillatory Gilbert damping as a function of the NiFe thickness in the Nb/NiFe/Nb heterostructures measured at $T$ = 10, 7, 5, and 4 K, respectively. The oscillating feature below $T_C$ ($T$ = 4 and 5 K) is dramatically enhanced compared to that above $T_C$ ($T$ = 10 K).

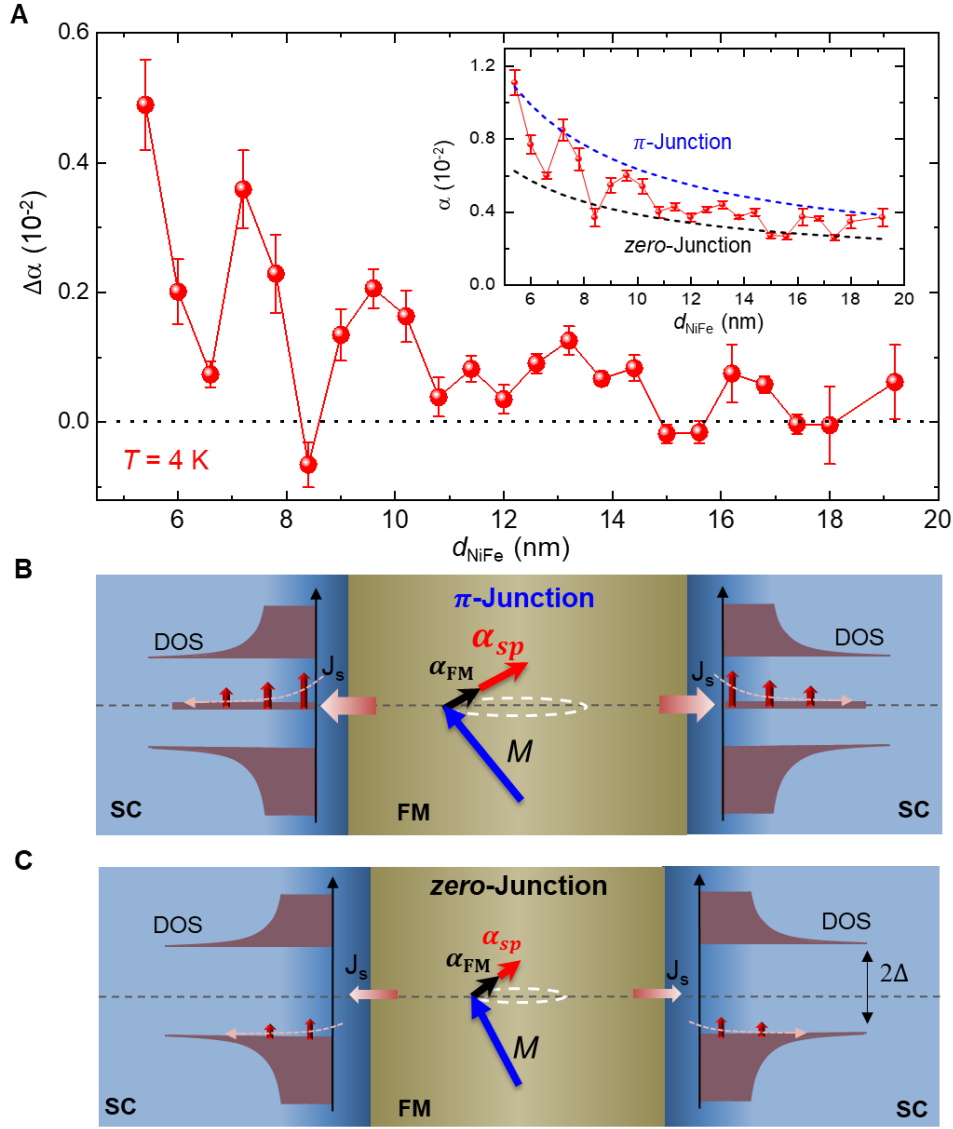

**Fig. 4. Physical mechanism of the giant oscillatory Gilbert damping in Nb/NiFe/Nb junctions.**
(**A**) The NiFe thickness dependence of the Gilbert damping difference ($\Delta\alpha$) between the Nb/NiFe/Nb $\pi$- and *zero*-junctions at $T = 4$ K. Inset: The Gilbert damping of *zero*- and $\pi$-junctions. The solid balls represent the experimental data, the blue and black dash lines are the guide lines for $\pi$- and *zero*-junctions, respectively. For both guide lines, the damping is expected to behave as $\alpha \sim 1/d_{NiFe}$. (**B-C**) Illustration of the spin pumping via the ABS and the enhanced Gilbert damping for Nb/NiFe/Nb $\pi$- and *zero*-junctions, respectively. The red thick-arrows indicate the pumping and relaxation of the spin current in SCs.

# Supplementary Materials for

## Giant oscillatory Gilbert damping in superconductor/ferromagnet/superconductor junctions

**Authors**

Yunyan Yao[1,2][†], Ranran Cai[1,2][†], Tao Yu[3], Yang Ma[1,2], Wenyu Xing[1,2], Yuan Ji[1,2], Xin-Cheng Xie[1,2,4,5], See-Hun Yang[6]*, and Wei Han[1,2]*

**This SM file includes:**

- Supplementary Materials and Methods
- fig. S1. The crystalline properties of the Nb/NiFe/Nb heterostructures.
- fig. S2. Gilbert damping measurement of Nb/NiFe/Nb heterostructures at $T = 10$ K.
- fig. S3. NiFe thickness dependence of Gilbert damping at $T = 50$ K.
- fig. S4. NiFe thickness dependence of $T_C$ for the Nb/NiFe/Nb heterostructures.
- fig. S5. Measurement of the Gilbert damping of Nb/NiFe/Nb heterostructures at $T = 4$ K.
- fig. S6. The effect of FMR measurement on the $T_C$ of Nb/NiFe/Nb heterostructures.
- fig. S7. The measurement of Josephson coupling in Nb/NiFe/Nb junctions.
- fig. S8. Illustration of magnetization dynamics and spin pumping in the SC/FM/SC heterostructures due to Bogoliubov quasiparticles and equal spin-triplet Cooper pairs.
- fig. S9. Calculation of the enhanced Gilbert damping due to spin pumping via the ABS at $T = 4$ K.
- fig. S10. Gilbert damping of control sample of bilayer Nb/NiFe junctions.
- fig. S11. Thickness dependence of gyromagnetic ratio, effective magnetization and inhomogeneous half-linewidth.

## Supplementary Materials and Methods

### Section 1: Model of oscillating Gilbert damping above $T_C$.

The oscillatory Gilbert damping in normal metal (NM)/ferromagnet (FM)/NM heterostructures arising from quantum interference effect is analyzed based on previous theory by Mills (36). Within the linear response theory, the enhanced Gilbert damping is related to the dynamical spin susceptibility ($\chi^{-+}(\Omega)$) of conduction electrons in a FM,

$$\alpha_{sp} = \frac{J^2 M_s V}{2N^2 \hbar^3 \gamma} \Lambda_2 \tag{S1}$$

where $\Lambda_2 = \text{Im}\left(\frac{d\chi^{-+}(\Omega)}{d\Omega}\bigg|_{\Omega=0}\right)$. Using one dimensional model, we obtain

$$\Lambda_2 = \frac{1}{\pi^2} \int_{FM} dx dx' \text{Im}[G_\uparrow(x, x', \epsilon_F)] \text{Im}[G_\downarrow(x, x', \epsilon_F)] \tag{S2}$$

where $G_\sigma(x, x', \epsilon_F)$ is the Green's function for conduction electrons with $\sigma$-spin at the Fermi energy ($\epsilon_F$). In a FM, $G_\sigma(x, x', \epsilon_F)$ is related to the exchange energy.

$$\left[-\frac{\hbar^2}{2m}\frac{d}{dx^2} - \epsilon \pm E_{ex}\right] G_\sigma(x, x', \epsilon_F) = \delta(x - x') \tag{S3}$$

For the FM film with a thickness $d_{FM}$ in $-d_{FM}/2 < x < d_{FM}/2$, the Green's function satisfies the relation

$$G_\sigma(x, x', \epsilon_F) = G_\sigma(x', x, \epsilon_F) = G_\sigma(-x, -x', \epsilon_F) \tag{S4}$$

Hence, the imaginary part of the Green's function could be expressed by

$$\text{Im}[G_\sigma(x, x', \epsilon_F)] = -\pi\{N_{F\sigma}\cos[k_{F\sigma}(x-x')] + N'_{F\sigma}\cos[k_{F\sigma}(x+x')]\} \tag{S5}$$

where $k_{F\sigma} = \sqrt{\frac{2m}{\hbar^2}(\epsilon_F \mp E_{ex})}$ is the Fermi wave-vector in the FM, $N_{F\sigma}$ and $N'_{F\sigma}$ are equivalent to the density of states and the modulation amplitude of the local density of states, respectively. For the same position of $x$, the local density of states is equal to

$$N_\sigma(x, \epsilon_F) = N_{F\sigma} + N'_{F\sigma} \cos[2k_{F\sigma}x] \tag{S6}$$

Since $E_{ex}$ is much smaller compared to $\epsilon_F$, the spatial modulation of the local density of states is negligible. The combination of equations (S2) and (S5) leads to

$$\Lambda_2 = \int_{-d_{FM}/2}^{d_{FM}/2} dx dx' \{N_{F\uparrow} \cos[k_{F\uparrow}(x-x')]\} * \{N_{F\downarrow} \cos[k_{F\downarrow}(x-x')]\} \tag{S7}$$

$$= 2N_{F\uparrow}N_{F\downarrow}\left\{\frac{1}{(k_{F\uparrow} - k_{F\downarrow})^2}\sin^2\left[\frac{k_{F\uparrow} - k_{F\downarrow}}{2}d_{FM}\right] + \frac{1}{(k_{F\uparrow} + k_{F\downarrow})^2}\sin^2\left[\frac{k_{F\uparrow} + k_{F\downarrow}}{2}d_{FM}\right]\right\}$$

Clearly, the enhanced Gilbert damping is expected to oscillate as a function of the FM thickness with two periods of $2\pi/[k_{F\uparrow} - k_{F\downarrow}]$ and $2\pi/[k_{F\uparrow} + k_{F\downarrow}]$. For real FM materials, such as NiFe with $(k_{F\uparrow} + k_{F\downarrow}) \gg (k_{F\uparrow} - k_{F\downarrow})$, the second term in the equation (S7) could be negligible, leaving only one oscillating period of $2\pi/[k_{F\uparrow} - k_{F\downarrow}]$. When the FM thickness is equal to $2n\pi/[k_{F\uparrow} - k_{F\downarrow}]$, a lower Gilbert damping is obtained. On the other hands with FM thickness of $(2n + 1)\pi/[k_{F\uparrow} - k_{F\downarrow}]$, a larger Gilbert damping is obtained.

**Section 2: Calculation of the enhanced Gilbert damping in Nb/NiFe/Nb by spin pumping via Andreev bound states (ABS).**

As the reciprocal process of the spin transfer torque, conventional spin pumping is achieved by the magnetization torques provided by the driven quasiparticle carriers (*29, 52-54*), which are the electrons in the normal metals. In SC/FM heterostructures, however, the quasiparticle carriers can be either Bogoliubov quasiparticles or ABS (*42*), which lie above and within the superconducting gaps, respectively. Therefore, it is desirable to formulate and estimate the contribution to the spin pumping via the ABS (*55*), when the temperature is much smaller than the superconducting critical temperature.

Without loss of generality, we start the analysis from a left-propagating electron of energy $\varepsilon$ and spin $\sigma = \{\uparrow, \downarrow\} = \{+, -,\}$. When the Zeeman splitting *J* is much smaller than the Fermi energy $E_F$, it has momentum

$$k_\sigma = k_F + (\varepsilon + \sigma J)/(\hbar v_F), \tag{S8}$$

where $v_F$ is the Fermi velocity of the electron. When one electron goes from the FM to the SCs, it is reflected as a hole by the Andreev reflection at the right FM/SC interface; this hole has a phase shift $\chi = -\arccos(\varepsilon/\Delta)$ with respect to the electron (*56*), where $\Delta$ is the superconducting gap. Similarly, when a hole goes from the metal to the superconductor at the left FM/SC interface, an electron can be reflected. With a proper energy, the Andreev reflections can form a closed path, as a result of which the ABS forms. This requires that the phase accumulated in the reflections satisfies the Sommerfeld quantization condition, i.e. in the ballistic regime,

$$\frac{\varepsilon L}{\hbar v_F} + \sigma \frac{J d_{NiFe}}{\hbar v_F} - \arccos\left(\frac{\varepsilon}{\Delta}\right) = n\pi + \frac{\varphi}{2}, \tag{S9}$$

where $\varphi$ is the phase difference between the two superconductors, $d_{NiFe}$ is the thickness of the FM layer and $n$ is an integer. Since $\hbar v_F/\Delta \geq 100$ nm with $v_F = 2.2 \times 10^5$ m/s and $\Delta = 1$ meV at $T = 4$ K in our experiment (*17, 57, 58*), $d_{NiFe} < 19$ nm $<< \hbar v_F/\Delta$ such that the first term in Eq. (S9) can be safely disregarded. For the FMR measurements with open-circuited configuration, the junctions always stay in the ground states (*43-45*). For $\pi$-junctions, there is a $\pi$-phase shift in the current-phase relationship curves compared to *zero*-junctions, i.e., the properties of $\varphi = 0$ of a $\pi$-junction is the same as those of $\varphi = \pi$ of a *zero*-junction. Since this $\pi$-phase shift is already taken into account by the FM exchange field, the ABS energy of the $\pi$-junctions can be obtained at $\varphi = 0$ in the ground states, which is similar to that of $\varphi = \pi$ of *zero*-junctions. Hence, the energy of the ABS can be described by $\varepsilon_0 = \pm\Delta\cos\left(\frac{J d_{NiFe}}{\hbar v_F}\right)$ for ideal case with perfect transparency of electrons/holes.

In reality, the interfacial scattering and transport conditions (ballistic or diffusive regimes) of FM could affect the energy of the ABS. Following previous studies (*42, 59*), a transmission coefficient (*D*) could be introduced to describe this issue, which is close to unity in the ballistic regime but can also be large in the diffusive regime with an ideal transparency at the interface (*43, 60, 61*). In this work, we focus on the ideal cases with perfect transparency of electrons/holes. The energy of the ABS oscillates from the edge of the superconducting gap to the zero-energy with respect to the FM thickness (fig. S9A).

The pumped spin current reads (*29, 52-54*),

$$\mathbf{J}_s(t) = \frac{\hbar}{4\pi} g_{eff}^{\uparrow\downarrow} \mathbf{m} \times \frac{d\mathbf{m}}{dt}, \tag{S10}$$

where $\mathbf{m}$ is the magnetization unit vector, and we define the effective mixing spin conductivity $g_{eff}^{\uparrow\downarrow}$ at the finite temperature via the zero-temperature one $g^{\uparrow\downarrow}$ by (*53, 55*)

$$g_{eff}^{\uparrow\downarrow} = n_0 \int d\varepsilon \frac{df(\varepsilon)}{d\varepsilon} \text{Re}[g^{\uparrow\downarrow}(\varepsilon)]. \tag{S11}$$

Here, $n_0$ is the number of the conduction channel that roughly corresponds the conduction electron density at the interface and $f(\varepsilon) = 1/\{\exp[\varepsilon/(k_B T)] + 1\}$ is the Fermi-Dirac distribution of electron at the temperature *T*. Importantly, in the ballistic limit $\text{Re}[g^{\uparrow\downarrow}(\varepsilon)] = 1$ when $\varepsilon = \varepsilon_0$; it has width $\Delta\varepsilon$ depending on the FM thickness $d_{\text{NiFe}}$ in the ballistic regime or the mean free path $l_m$ in the diffusive regime. By the uncertainty principle, $\Delta\varepsilon \Delta t = 2\pi\hbar$, where $\Delta t = l_m/v_F$ is the propagation time of the electron in the junction, leading to $\Delta\varepsilon \sim 2\pi\hbar v_F/l_m$. By further considering the degeneracy due to spin ($\times 2$) and the existence of two interfaces ($\times 2$), we thus can estimate

$$g_{\text{eff}}^{\uparrow\downarrow} \sim 8\pi n_0 \frac{\hbar v_F}{l_m} \frac{df(\varepsilon_0)}{d\varepsilon}. \tag{S12}$$

The pumped spin current carries the angular momentum away from the precessing magnetization and hence cause an enhanced Gilbert damping, which is described by

$$\delta\alpha = 2\gamma \frac{\hbar^2 v_F}{M_s l_m d_{\text{NiFe}}} \frac{df(\varepsilon_0)}{d\varepsilon}, \tag{S13}$$

where $\gamma$ is electron gyromagnetic ratio and $M_s$ is the saturated magnetization of the ferromagnet.

We are now ready to estimate the contribution of ABS to the Gilbert damping at $T = 4$ K with varying transmission coefficient. We take $n_0 = 0.5 \times 10^{16}$ m$^{-2}$ following Ref. 44, $l_m \sim 3$ nm, $v_F = 2.2 \times 10^5$ m/s, $J = 400$ meV and $\mu_0 M_s \approx 1$ T from previous experimental results (*17*). With superconducting gaps $\Delta \approx 1$ meV at $T = 4$ K for Nb (*57, 58*), Fig. S8A plots the normalized energy of ABS by the superconducting gap at $T = 4$ K as a function of $d_{\text{NiFe}}$ for the ideal transparency case. The oscillation of the Gilbert damping can be resolved by using the FM exchange field-induced phase shift of $\frac{J d_{\text{NiFe}}}{\hbar v_F}$ (fig. S9B). For simplicity, we have disregarded the possible thickness dependence of the superconducting gaps and magnetizations. To be noted, our theoretical estimation is based on a simplified model that assumes $D = 1$. For the diffusive regime or the case of non-perfect transparency of electrons at the interface (*42, 43*), similar oscillating behaviors of ABS (or DOS) in the SCs can also be preserved. For example, the oscillating ABS (or DOS) in the SCs have been shown to exist in the diffusive regime theoretically (*6, 46*), and indeed, the *zero* to $\pi$ transitions have been experimentally observed in both the ballistic and diffusive regimes from the supercurrent measurements (*11, 17*). To fully understand the experimental observation of the oscillatory Gilbert damping in the diffusive regime, further theoretical studies are needed.

**Section 3: Measurement of the Josephson coupling in Nb/NiFe/Nb.**

The Nb/NiFe/Nb Josephson devices are fabricated using the shadow mask techniques during the films growth. As shown in figs. S7A and S7B, the Josephson devices have a junction area ($A$) of ~ 80 μm × 80 μm, and the other areas are electrically isolated by a 100 nm AlO$_x$ layer. The Josephson current is measured by standard a.c. lock-in technique. The normalized differential resistances (dV/dI) measured on the Nb/NiFe (5 nm)/Nb junction at various temperatures are shown in fig. S7C. The critical current ($I_c$) is defined as point where the differential resistance increases above the value for the zero-bias current. The normal resistance ($R_n$) is determined to be the saturated value of the normal states of the Josephson coupling measurement. The measured area-resistance product ($R_nA$) of ~ $5 \times 10^{-10}$ $\Omega m^2$ is higher than that reported in metallic Josephson junction (*17, 61*), and comparable to that of FM Josephson junction with a thin tunnel barrier (*62*). This behavior indicates that there is more likely a thin NiFeO$_x$ layer (indicated by Fig. S8B) in the junction formed during the AlO$_x$ growth step in the presence of oxygen gas. As the temperature increases, $I_c$ and the characteristic voltage ($I_cR_n$) decrease (figs. S7C and S7D). Clear Josephson currents are observed on the Nb/NiFe (5 nm)/Nb junction and Nb/NiFe (10 nm)/Nb junction (figs. S7E and S78F). And the estimated SC gap energy is ~ 0.9 meV at $T$ = 2 K (*1, 43*), which is comparable to the value of ~1.36 meV at $T$ = 0 K estimated from $T_C$ of ~ 8.5 K (fig. S4). On the other hand, no Josephson current could be observed in the Nb/NiFe (30 nm)/Nb junction (figs. S7E and S7F). The absence of Josephson current in Nb/NiFe (30 nm)/Nb junction indicates that there is no long-range spin-triplet Josephson coupling in the Nb/NiFe/Nb heterostructures in our experiment.

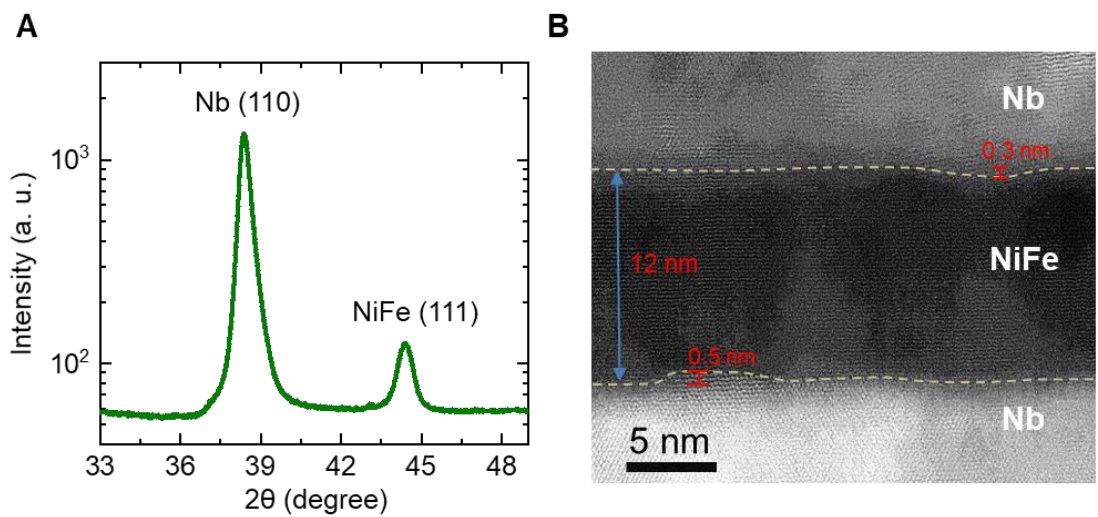

**fig. S1. The crystalline properties of the Nb/NiFe/Nb heterostructures.** (**A**) The θ-2θ X-ray diffraction results measured on the typical Nb/NiFe (12 nm)/Nb sample, where Nb (110) and NiFe (111) peaks are observed. (**B**) High-resolution transmission electron micrographs measured on the typical Nb/NiFe (12 nm)/Nb sample. The dashed lines show the interfaces between Nb and NiFe layers. The red bars indicate the deviation of NiFe at the interface.

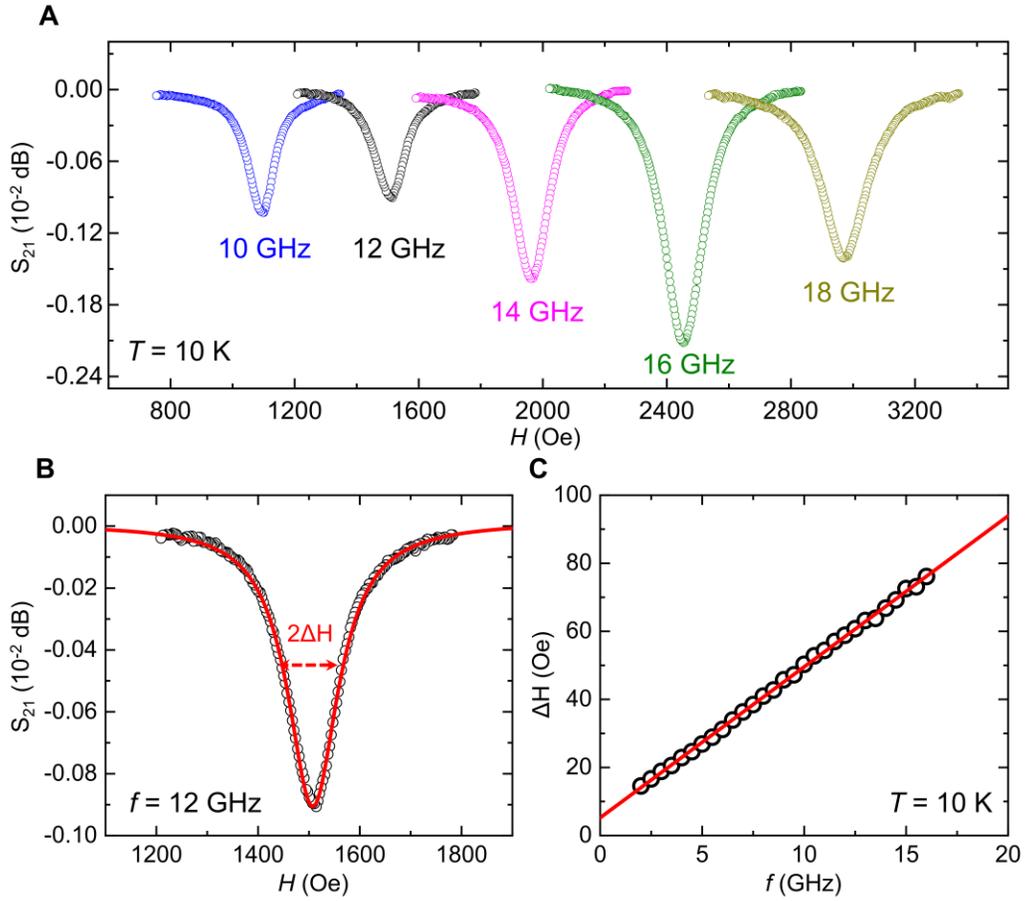

**fig. S2. Gilbert damping measurement of Nb/NiFe/Nb heterostructures at $T$ = 10 K.** (**A**) The typical FMR spectra as a function of magnetic field with microwave frequency (*f*) of 10, 12, 14, 16, and 18 GHz, respectively. (**B**) The typical FMR spectrum measured with $f$ = 12 GHz (black circles) and the Lorentz fitting curve (red line). ΔH is the half linewidth of the FMR signal. (**C**) The determination of the Gilbert damping from ΔH vs. *f*. The red line indicates the best linear-fitting curve. These results are obtained on the typical Nb/Py (12 nm)/Nb sample.

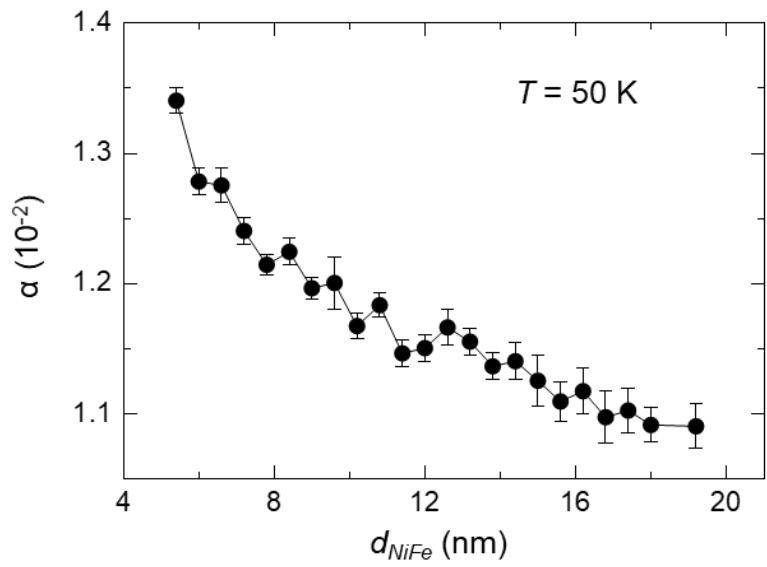

**fig. S3. NiFe thickness dependence of Gilbert damping at $T$ = 50 K.**

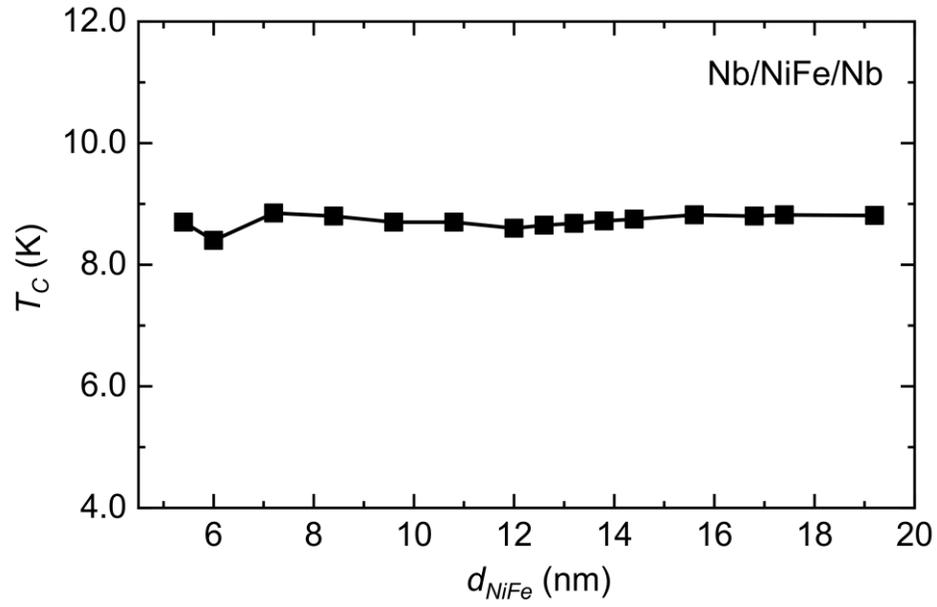

**fig. S4. NiFe thickness dependence of $T_C$ for the Nb/NiFe/Nb heterostructures.** The $T_C$ is determined from the zero-resistance temperature via four-probe resistance measurement.

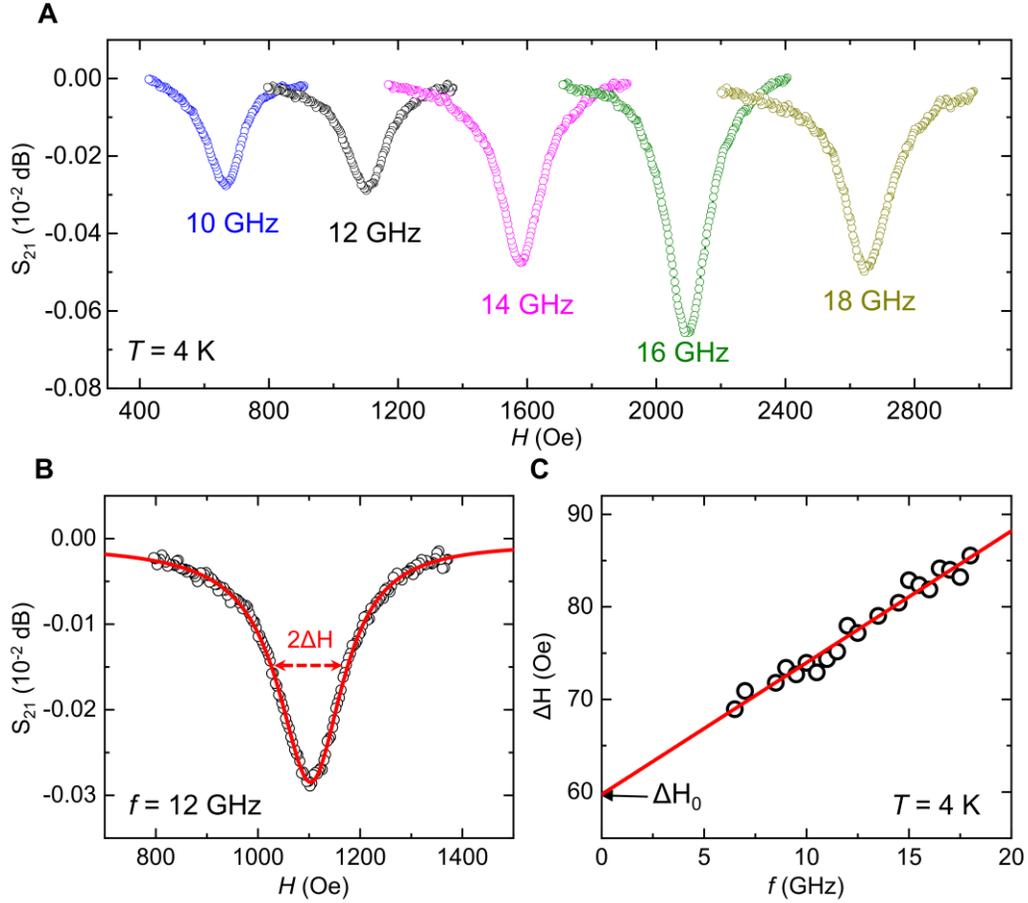

**fig. S5. Measurement of the Gilbert damping of Nb/NiFe/Nb heterostructures at $T$ = 4 K.** (**A**) The typical FMR spectra as a function of magnetic field with microwave frequency (*f*) of 10, 12, 14, 16, and 18 GHz, respectively. (**B**) The typical FMR spectrum measured with $f$ = 12 GHz (black circles) and the Lorentz fitting curve (red line). ΔH is the half linewidth of the FMR signal. (**C**) The determination of the Gilbert damping from ΔH vs. *f*. The red line indicates the best linear-fitting curve. These results are obtained on the typical Nb/Py (12 nm)/Nb sample.

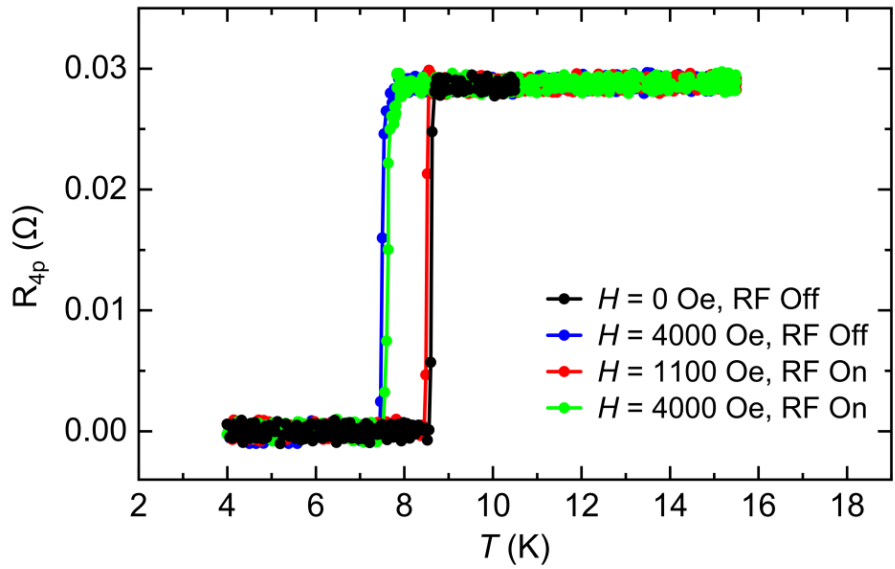

**fig. S6. The effect of FMR measurement on the $T_C$ of Nb/NiFe/Nb heterostructures.** The four-probe resistances *vs.* temperature are probed from the typical Nb/NiFe (12 nm)/Nb sample with/without the presence of the in-plane magnetic field and microwave power.

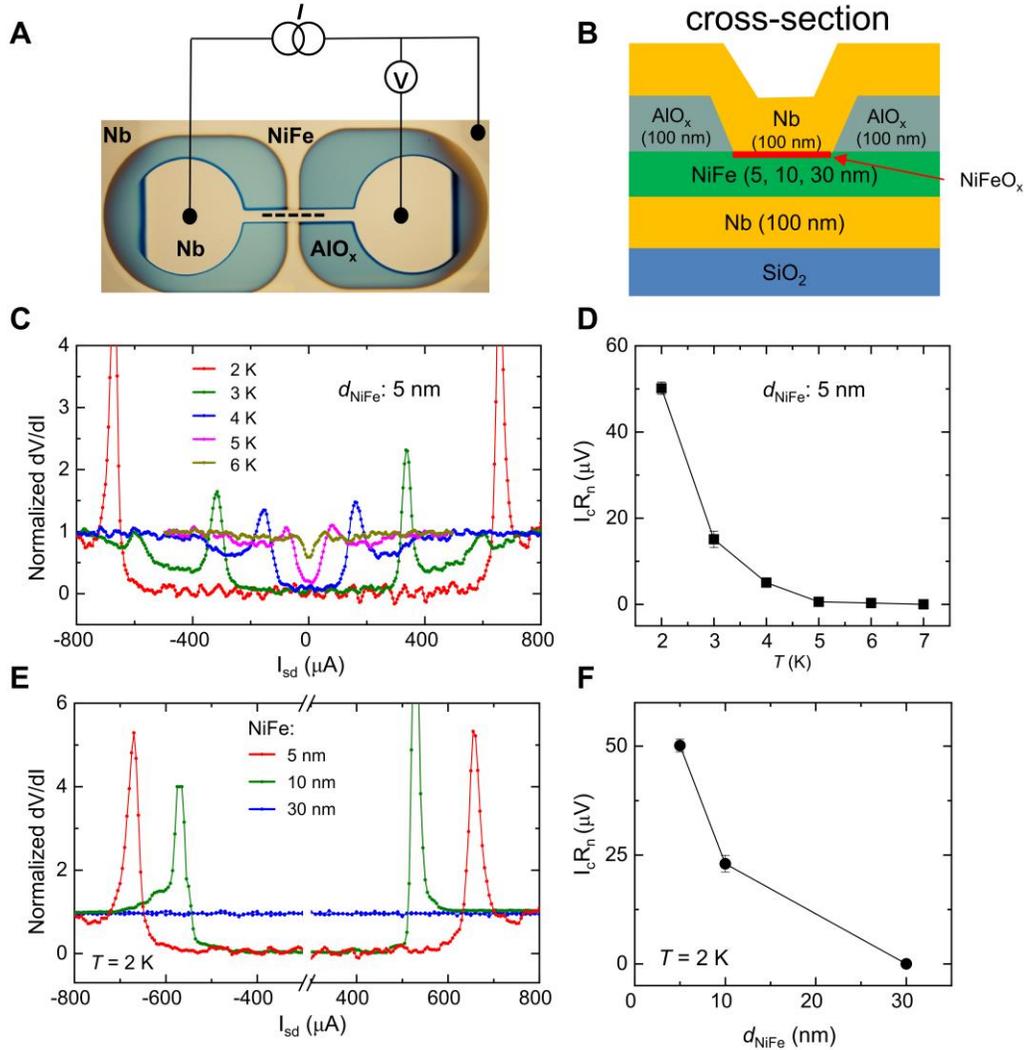

**fig. S7. The measurement of Josephson coupling in Nb/NiFe/Nb junctions.** (**A**) The optical image of a typical Nb/NiFe/Nb Josephson device and schematic of the electrical measurement geometry. (**B**) The cross-section of the Josephson devices with a junction area of ~ 80 μm × 80 μm. At the junction, a thin oxide layer of NiFeO$_x$ is mostly liked formed on the top surface of NiFe during the growth of AlO$_x$ in the presence of oxygen. (**C**) The normalized differential resistance (dV/dI) as a function of the bias current measured on the Nb/NiFe (5 nm)/Nb junction from $T$ = 2 to 6 K. (**D**) The temperature dependence of the characteristic voltage ($I_cR_n$) of the Nb/NiFe (5 nm)/Nb Josephson junction. (**E**) The normalized differential resistance as a function of the bias current of the Nb/NiFe/Nb junctions ($d_{NiFe}$ = 5, 10 and 30 nm) at $T$ = 2 K. (**F**) The NiFe thickness dependence of the characteristic voltages at $T$ = 2 K.

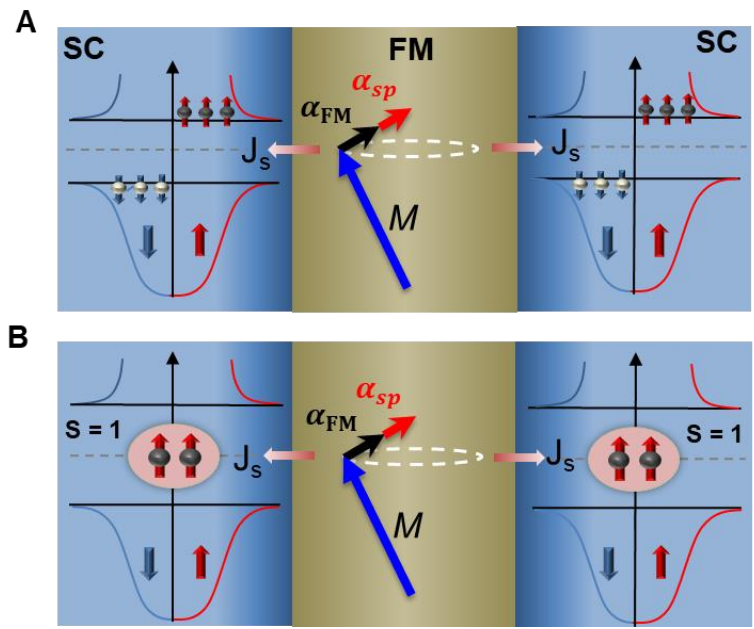

**fig. S8. Illustration of magnetization dynamics and spin pumping in the SC/FM/SC heterostructures due to Bogoliubov quasiparticles (A) and equal spin-triplet Cooper pairs (B).** The dark and light balls represent the electron-like and hole-like quasiparticles respectively. The red and blue arrows indicate the spin up and spin down respectively.

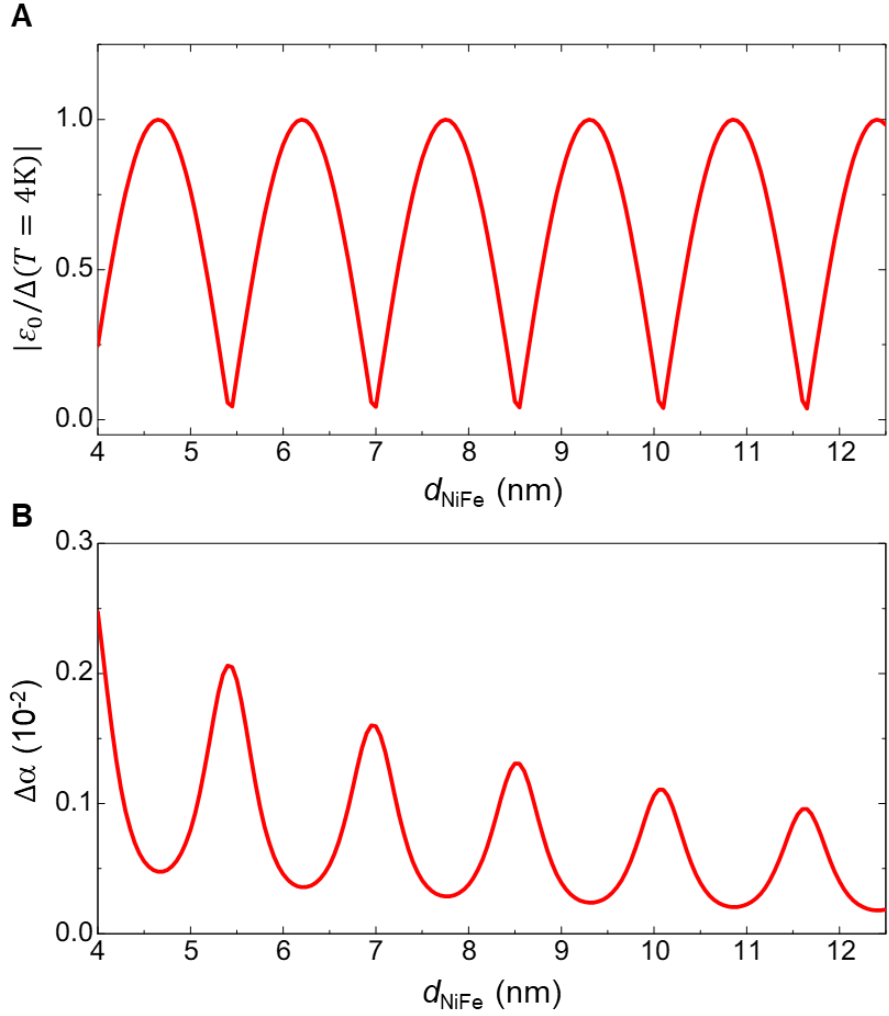

**fig. S9. Calculation of the enhanced Gilbert damping due to spin pumping via the ABS at $T = 4$ K.** (**A**) The normalized energy of ABS by the superconducting gap at $T = 4$ K as a function of $d_{NiFe}$ for the ideal transparency case. (**B**) The enhanced Gilbert damping via ABS as a function of $d_{NiFe}$.

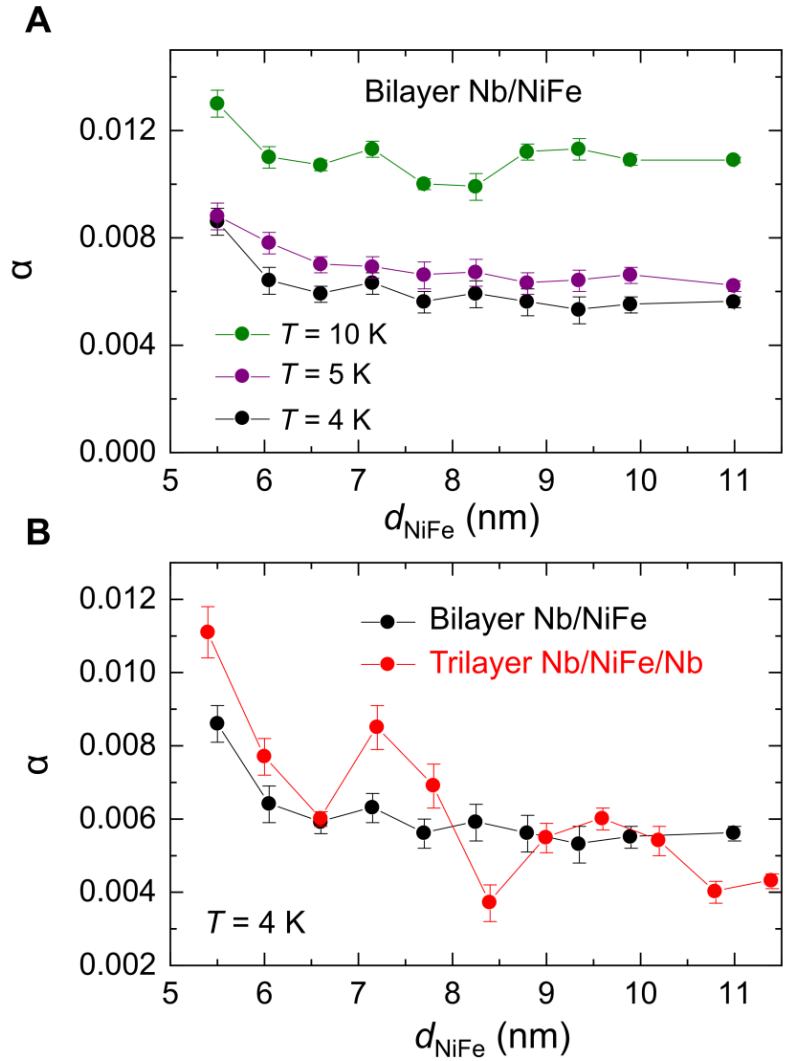

**fig. S10. Gilbert damping of control sample of bilayer Nb/NiFe junctions.** (**A**) Gilbert damping of bilayer Nb/NiFe junctions at $T$ = 4, 5, and 10 K. (**B**) Comparison of the Gilbert damping of bilayer Nb/NiFe and trilayer Nb/NiFe/Nb junctions at $T$ = 4 K.

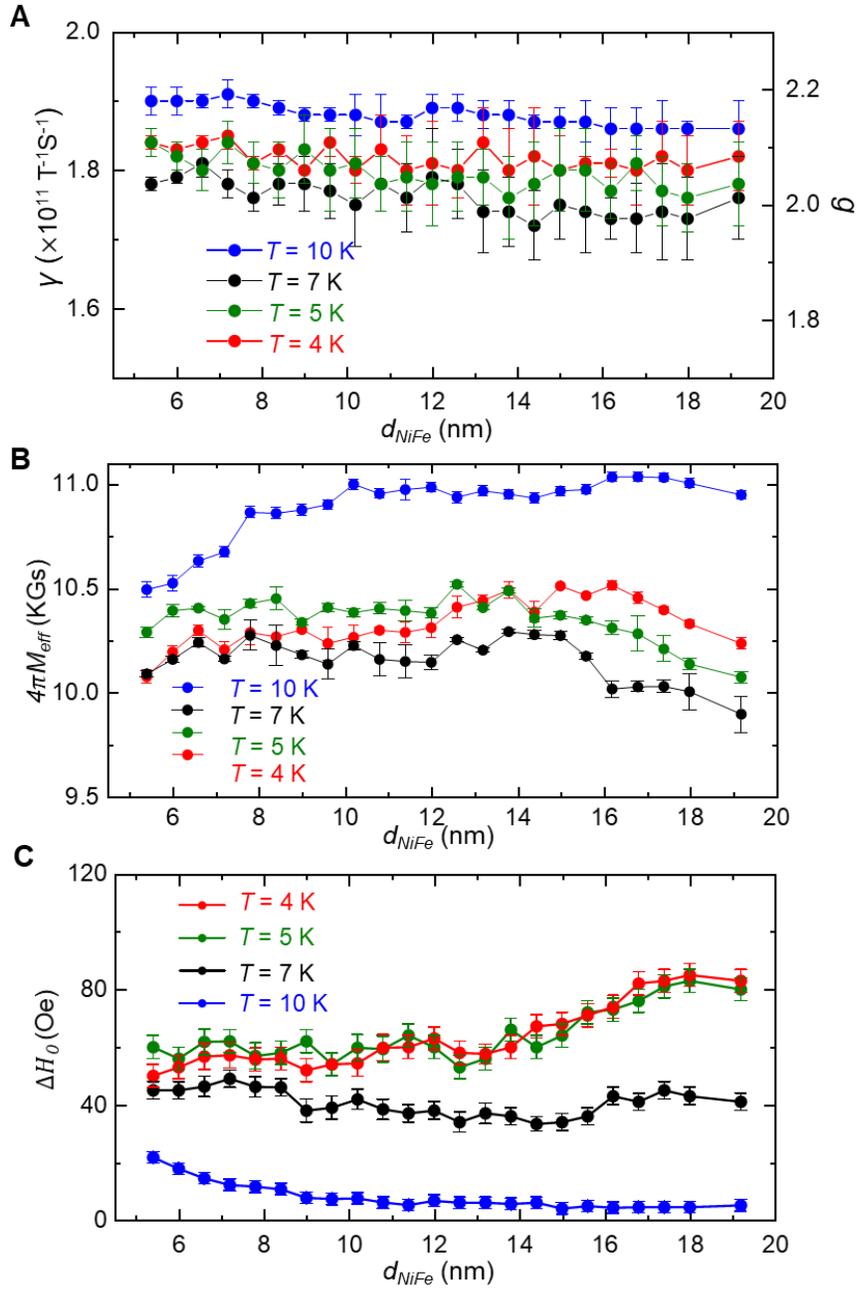

**fig. S11. Thickness dependence of gyromagnetic ratio (and g factor) (A), effective magnetization (B) and inhomogeneous half-linewidth (C).** The blue, black, green and red dotted-lines represent to temperature of $T$ = 10, 7, 5 and 4 K, respectively.